\documentstyle[12pt,epsf,a4]{article}
\begin{document}
\pagestyle{plain}
\parsep  6pt plus 1pt minus 1pt
\parindent 12truemm
\def\beq{\begin{equation}}
\def\eeq{\end{equation}}
\newcommand{\bea}{\begin{eqnarray}}
\newcommand{\eea}{\end{eqnarray}}
\newcommand{\nn}{\nonumber}
\def\simleq{\; \raise0.3ex\hbox{$<$\kern-0.75em
      \raise-1.1ex\hbox{$\sim$}}\; }
\def\simgeq{\; \raise0.3ex\hbox{$>$\kern-0.75em
      \raise-1.1ex\hbox{$\sim$}}\; }
\def\noi{\noindent}
\font\boldgreek=cmmib10
\textfont9=\boldgreek
\mathchardef\myrho="091A
\def\bfrho{{\fam=9 \myrho}\fam=1}
\def\R{ {\rm R \kern -.31cm I \kern .15cm}}
\def\C{ {\rm C \kern -.15cm \vrule width.5pt
\kern .12cm}}
\def\Z{ {\rm Z \kern -.27cm \angle \kern .02cm}}
\def\N{ {\rm N \kern -.26cm \vrule width.4pt \kern .10cm}}
\def\1{{\rm 1\mskip-4.5mu l} }

\vbox to 2 truecm {}
\centerline{\large \bf The controversy about "$1/m_Q$ duality violation" : } 
\centerline{\large \bf a quark model point of view}
\par \vskip 10 truemm  
 
\centerline{A. Le Yaouanc$^{a}$, D. Melikhov$^{b,}$\footnote{On leave from {\it Nuclear Physics Institute, Moscow State University, Moscow, Russia}}, 
V. Mor\'enas$^{c}$, L. Oliver$^{a}$, O. P\`ene$^{a}$ and J.-C. Raynal$^{a}$} \par \vskip 4 truemm

$^a$ Laboratoire de Physique Th\'eorique, Universit\'e de Paris XI, \par 
B\^atiment 210, 91405 Orsay Cedex, France\footnote{
Unit\'e Mixte de Recherche - CNRS - UMR 8627}\\

$^b$ Institut f\"ur Theoretische Physik, Universit\"at Heidelberg, \par
Philosophenweg 16, D-69120, Heidelberg, Germany\\

$^c$ Laboratoire de Physique Corpusculaire, Universit\'e Blaise Pascal - CNRS/IN2P3,\par 63000
Aubi\`ere Cedex, France
\par \vskip 1 truecm

{\bf Abstract}. Non relativistic quark models have been invoked to support the statement of a $1/m_Q$ duality violation in semileptonic $B$ decays. However, as we recall, we have very explicitly shown that no $1/m_Q$ duality violation is present in totally integrated rates, in such quark models. Then : 1) it is shown that some contributions under discussion are misleadingly described as "$1/m_Q$ duality violation" ; as they stand, they are in fact parametrically much smaller :  they are ${\cal O}((1/m_Q)^{5/2})$ when properly referred to the total rate under discussion ; numerically this is below $10^{-2}$ ; 2) moreover, relying on our rigourous expansion of the harmonic oscillator model around the Shifman-Voloshin limit, it is shown that even such small terms are not present at all in the total rate, and must therefore merely cancel ; 3) finally, for physical masses, by an exact treatment of this particular, but not too unrealistic model, one finds a departure from free-quark decay rate only at the $10^{-3}$ level ; moreover, this departure is {\bf totally explained by the OPE}  to a high precision (with an error of $10^{-5}$ only), provided we include the $1/(m_Q)^3$ terms. Further terms are strongly suppressed. On the whole, from the various detailed studies which have been made, it appears highly probable that equality between inclusive decay and OPE holds very well for total decay rates in QCD, even for a realistic spectrum and kinematics, provided the OPE expansion itself is pushed sufficiently far. In other words, not only $1/m_Q$ duality violation cannot be present, but possible duality violation, at high orders, is small.\par 

\vskip 1 truecm

\section{Setting the controversy} The Minnesota group \cite{minnesota} has established and discussed in detail the question of duality in QCD \cite{minnesota}, as well as in the $QCD_2$ model \cite{lebed} . In particular, they have extensively argued the famous result that there is no $1/m_Q$ term in the ratio of inclusive to free quark total decay rate, in semi-leptonic decays of heavy quarks, a result that was already presented in ref. \cite{chay}. This statement derives from the OPE. {\bf "Duality" means equality of inclusive rate with OPE}, and, in this OPE, no $1/m_Q$ term is present. Note therefore that, to this order included, the inclusive rate should equate the free quark one. On the contrary, in ref.\cite{isgur}, Nathan Isgur has pointed out effects presented as a "$1/m_Q$ duality violation" , a statement which has been often referred to.
This would hold in quark models and extend presumably to QCD, since it has a very general structure. 

In the papers \cite{articleNR,lettreNR,OH}, we have studied the question for totally inclusive rates, using the
same type of hadronic models as used by Isgur, namely quark models ; more precisely, we use rigourously treated, strictly non relativistic quark models, 
which have the advantage that one can calculate directly the inclusive rates as sums over bound states. Then, {\bf we do not find duality violation at this order $1/m_Q$} in the {\bf integrated} rates, i.e. the rate is found to be in accordance with the OPE expectation. An important fact related to this conclusion is that 
the Bjorken and Voloshin sum rules are automatically satisfied in this model. We have explained here where our disagreement with ref.\cite{isgur} is precisely
taking place in ref. \cite{CKM/hep-ph/0304132} . 

The possibility suggested in ref.\cite{isgur} is continuing to be invoked in several experimental or theoretical papers in spite of the many proofs which have been provided to the contrary, and although no indication has been found of such an effect, as admitted in these papers
themselves. Then, we think it worth returning to the question in detail. This paper extends our contribution to the CERN Yellow book \cite{CKM/hep-ph/0304132}. The present paper contains much more complete discussions. In particular, additional informations on complementary aspects often intervening in the discussions are included (sections \ref{channels} and \ref{numerical}).

\section{$1/m_Q$ expansions in the non relativistic quark model}
For sake of simplicity, we present the results for the rates as well the subsequent calculations in the rest frame of the decaying particle, although we have also done the calculations for a moving one.
We sum up our conclusions from the model {\bf calculations} in the formula :
\bea
R_{sl}={\Gamma_{inclusive} \over \Gamma_{free~quark}}=
1+{\cal O }\left({1 \over m_b^2}\right)+{\cal O }\left({1 \over m_b^2} {\Delta \over \delta m_Q} \right)+...\label{R}
\eea
valid for sufficiently regular potentials. 

$\Delta$ is a typical excitation energy for the mesons levels ; $\delta m_Q=m_b-m_c$. $\Gamma_{inclusive}$ and its expansion have been calculated as the sum of the widths of any possible state. One can see that the terms that are added to $1$, correcting the free-quark rate, are corresponding to the expected terms of the OPE or more precisely, of the {\it short time} expansion which plays the role of OPE in this type of models \footnote{The operators in the series are local in time, but, unlike in QCD, and not surprisingly for a potential theory, their are not local in space. Note that we retain in the paper the term "OPE",  for shortness, but also to underline the parallelism with QCD.}; this corresponds to the general statement of duality, because duality means basically that the inclusive rate is given by an OPE expansion, beginning with the free-quark rate. Therefore, we find duality to be satisfied explicitly in the models.  Note that there are two small parameters in the OPE expansion, one in addition to $\Lambda_{QCD}/m_b$ : we can choose as the second one ${\Lambda_{QCD} \over \delta m_Q}$. 

To present a simpler expansion, we have used above the Shifman-Voloshin (S-V) limit \cite{SV}, where, in QCD, $\Lambda_{QCD}\ll\delta m_Q =m_b-m_c\ll m_b$. The second small parameter ${\Lambda_{QCD} \over \delta m_Q}$ is then
large with respect to $\Lambda_{QCD}/m_b$. In the non relativistic model, the role of $\Lambda_{QCD}$ with a light current quark mass is traded for $\Delta $, the excitation energy with respect to the ground state, and the light constituent quark mass $m_d$.   

The expansion begins with ${1 \over m_b^2}$. There is no ${1 \over m_b}$ term.
This corresponds in the OPE to the fact that we do not find operators with a ${1 \over m_b}$ coefficient. Moreover, there are no terms with a positive power $n$ of $\delta m_Q$, i.e. terms like ${(\delta m_Q)^n \over m_b^2}$ are not present. Explicit calculations of the various bound state contributions to $\Gamma_{inclusive}$ in this model show that terms of order ${\cal O }({ (\delta m_Q)^2 \over m_b^2})$ (they correspond to form factor effects) and the smaller ${\cal O}({\delta m_Q \over m_b^2})$ appear in the exclusive rates, but they cancel in the sum. The respective cancellations are ensured respectively by the Bjorken sum rule for ${\cal O }({(\delta m_Q)^2 \over m_b^2})$ terms and, as we have noticed, by the Voloshin sum rule for ${\cal O }({\delta m_Q \over m_b^2})$ terms  \cite{QCD}. It is found that these sum rules are automatically satisfied in the model, which explains the cancellation of the two type of terms. 

In the usual $1/m_Q$ expansion where $m_c/m_b$ is held fixed, one has instead $\delta m_Q \propto m_Q$, 
and ${\cal O }({ (\delta m_Q)^2 \over m_b^2})$ and ${\cal O}({\delta m_Q \over m_b^2})$ terms will be counted respectively as $(1/m_Q)^0$
 and $(1/m_Q)$. Since the first type of terms correspond to form factor effects, i.e. bound state effects, they are naturally not suppressed by the heavy quark mass in individual state contributions, but they must disappear in the total rate, so as to get the free quark rate at leading order. On the other hand, the fact that terms of order ${\cal O }({\delta m_Q \over m_b^2})$, which would correspond to order $1/m_Q$ are absent in the ratio $R_{sl}$, may be expressed as "absence of $1/m_Q$ duality violation", because it is in agreement with OPE. As we see, the surviving terms, written in eq. (\ref{R}), are suppressed with respect to the above ${\cal O }({\delta m_Q \over m_b^2})$, by powers of $1/m_b$ or by powers of $\Delta/\delta m_Q$, $\Delta$ being the level spacing, of the order of $\Lambda_{QCD}$ , i.e. they are at most of order $(1/m_Q)^2$. 

\section{A completely explicit calculation} \label{exact} To confirm the above statements, it is useful to exhibit the phenomena of concern in a completely explicit model. In \cite{OH}, we have analysed very explicitly the cancellations which occur with an harmonic oscillator (H.O.) potential to produce these results. The interest of this model is that the truncation of states to the first band of orbital excitations (lowest $D^{**}$) becomes exact to the relevant order $1/m_b^2$ included, which allows to present a very simple analysis ; also it permits to perform {\bf completely explicit analytical calculations and expansions to high orders or numerical calculations to any accuracy}. In addition, it is close to the models 
which are invoked to support the arguments of \cite{isgur}, so that one can check with precision the various statements 
made in this paper. In the following $m_d$ is the light quark mass, $R^2$, the square of harmonic oscillator 
radius. We find, with $\Delta={1 \over m_d R^2}$ in the model (this is the difference between the ground state and the first excited level), and with a constant, momentum independent 
leptonic interaction\footnote{In the paper \cite{OH}, we have given the results for a more realistic 
leptonic interaction. The present interaction is chosen to fit as much as possible the two-body model of \cite{isgur}, with a remaining difference as regards parameters, which is explained at the end of section \ref{exact}} 
\bea
R_{sl}=1+{3 \over R^2 m_b^2}\left({1 \over 4}-{\Delta \over \delta m_Q}\right)+smaller~terms \label{ROH}
\eea
As we said, this expansion can be constructed through an OPE; for example, the first term in the bracket corresponds to the kinetic energy operator. 
In fact we have demonstrated that for {\bf regular} potentials as the H.O. model, the whole OPE series equates the exact total width up to terms smaller than any negative power \cite{lettreNR}. The conclusion to be drawn from eq.(\ref{ROH}) is once more that $1/m_Q$ duality violation, which would show up as a larger ${\cal O}({\delta m_Q \over m_b^2})$ term, is not present.
It remains then to explain where opposite conclusions invoking the work  \cite{isgur} may fail. We try to give an answer in sections \ref{real},\ref{cancellation}.

\section{The condition that many channels be open ($\delta m_Q >>\Delta$ )} \label{channels}
But before proposing this answer, we think useful devote a section to an
important question about the conditions of validity of the above expansions. Since doubts have indeed arisen from apparently contradictory statements, it is useful to discuss in some detail why we say that the ratio $\Delta/\delta m_Q$ must be small. The smallness of this parameter will play a crucial role in the following sections . This condition allows many states to be {\bf kinematically} allowed (which does not imply that transition amplitudes to all of them are large) ; this seems to be a natural condition for the OPE expansion, which requires inclusiveness, that is a large summation of states. However, the requirement of having many channels opened may seem to be contradicted by some well-known results; one may think for instance to the well known Shifman and Voloshin result that, in the SV limit, the free parton rate equals the sum of rates to ground state mesons $D^{(*)}$ only.
We show that the contradiction is only apparent. 

a) First, let us stress that the SV theorem does not imply that inelastic channels are absent ; what happens when many channels are open is that all the contributions, other than the one of the ground state, are small with respect to the latter : they include a factor $({\delta m_Q \over m_b})^2$ which is by definition small in the SV limit.

b) Second, one notes that the Shifman-Voloshin equality of the free quark rate and the elastic contribution is violated by a term of same order $({\delta m_Q \over m_b})^2$ -coming from the limited expansion
of the form factors. Although  $({\delta m_Q \over m_b})^2$ is small in the SV limit,
nevertheless, it is a term which is excluded in the totally inclusive rate. And this happens because it cancels precisely with the corresponding term from the inelastic contribution, leaving only terms small with  ${1 \over m_b^2}$, which are indeed allowed by eq. \ref{R}. 

c) Third, why and when is the the condition $\delta m_Q >>\Delta$ {\bf necessary} ? The answer is : it is in order that the first term in the expansion eq. \ref{R}, of the form ${1 \over m_b^2}$  dominates effectively over the next ones. If $\delta m_Q$ is not large with respect to $\Delta$, one has to take into account a whole series of term with increasing powers 
$({\Delta \over \delta m_Q})^n $.

More generally, there are admittedly several numerical situations where  the free quark rate is saturated very well "by a few channels ", although the mass is not heavy
or although not many channels are opened. In these latter cases,  there is neither an actual contradiction with our statement ; what happens is that the difference, although small, is no longer indicated by the first term in the above expansion, ${1 \over m_b^2}$ ; one cannot neglect the next terms in the $\Delta/\delta m_Q$ expansion ; the difference may be practically very small, but it comes out as the result of a cancellation of the various orders ;  it could be larger, or smaller, or of opposite
sign with respect to the first term . There is an example below, eq.(\ref{num}), in which we calculate the difference between the exact inclusive rate and the free quark decay rate in the realistic situation, where $m_b-m_c$ is not really small with respect to the two masses: then, one observes that the difference is small, but the first term, and not even the first two terms in the above S-V series are not sufficient to account for it. Other
examples, in $QCD_2$, given by Lebed and Uraltsev \cite{lebed}, exhibit similar phenomena. 

In short, it is when one desires to use the systematic expansions \ref{R} or \ref{ROH}, that {\bf $\Delta/\delta m_Q$ must be small}, i.e. that the number of open channels must be large. 
And this expansion is the only clearcut, well defined, statement.

\par \vskip 1 truecm

\section{Real order of magnitude of the alleged $1/m_Q$ duality violating terms} \label{real} Now, let us return to the initial question : what is the explanation of the apparent disagreement with the author of \cite{isgur}, who seems to have obtained  $1/m_Q$ duality
violating terms, in contradiction with our SV expansion, where such terms would show up as ${\delta m_Q \over m_b^2}$ terms and are not found in our calculation ? First, there seems to be a misunderstanding as to the magnitude of the effects considered in \cite{isgur}, induced by the expression "$1/m_Q$ duality
violation" repeatedly used, and which is very misleading (this section) : the terms are not of order ${\delta m_Q \over m_b^2}$, but smaller. Second, it seems that in addition very strong cancellations are operating in the total rate, which have been missed and lead to a complete disappearance of the terms (section \ref{cancellation}). We discuss these two points successively. We then proceed to an exact calculation which should hopefully discard any remaining doubt (section \ref{numerical}).\par

Let us explain our first point. In fact, in \cite{isgur}, it seems not to be actually disputed that the OPE expansion is basically right, and that the duality with free quark decay is satisfied within the expected accuracy in the region of phase space where the energy release is large $(t_{max}^{1/2}-t^{1/2})/\Delta >> 1$ (this condition means that many states
are kinematically allowed). This is certainly true in the physical situation when $t$ is small (i.e. large available three-momentum transfer). What may cause problem, according to \cite{isgur}, is only the region near $t_{max}$ where this condition is not satisfied. The main point of \cite{isgur} is that this could generate large effects of order $1/m_Q$. However, the expression "order $1/m_Q$" is confusing. According to us, the effects which are invoked are of relative order $1/m_Q$ only with respect to the decay {\bf over a small region of phase space} : the one where only the ground state can be produced. But -this is the first part of our objection-, these effects must be properly referred to the {\bf total decay rate}, since, in the present debate, it is the totally inclusive rate which is to be determined, (the aim is to determine $V_{cb}$ by comparison with the totally inclusive experimental rate);
the debate is about duality violation in the total rate ; then we find that the invoked effects are not of order $1/m_Q$  with respect to this totally inclusive free quark decay rate, but much smaller, by powers of $2 \Delta / \delta m_Q$, which amounts in the standard $1/m_Q$ expansion at fixed ratio of heavy masses, to further powers of the heavy mass because then, $\delta m_Q \propto m_b$. In fact, as we have emphasized above, $ \Delta / \delta m_Q$ must be taken as small in the general OPE expansion. Also, numerically, the terms are small realistically since  $2 \Delta / \delta m_Q$ is small (around $2~10^{-1}$, see the parameters below). This we show now.
 
Effect I. The first example given by Isgur is that the decrease of the ground state contribution with decreasing $t$ 
(i.e. increasing $|\vec{q}{\,}|$), due to the form factor, must be compensated by the increase of the excited 
states to maintain duality with free quarks. This is exactly guaranteed by the Bjorken sum rule in the heavy 
quark limit, but it is no more exact for finite mass, because there is a region below the $D^{**}$ threshold 
where only the ground state $D+D^{*}$ contribute.
Let us write quantitatively the term pointed in \cite{isgur}, with a constant leptonic interaction-the 
choice of this interaction is not crucial:
\bea
\delta \Gamma_I  \simeq- K {\rho^2 \over m_b^2 }
\int_{(\delta m_Q - \Delta)^2}^{(\delta m_Q)^2} dt |\vec{q}{\,}| |\vec{q}{\,}|^2 
\eea
where K is an irrelevant constant, $ |\vec{q}{\,}|^2 \simeq (\delta m_Q)^2 -t $ ; $-\rho^2 {|\vec{q}{\,}|^2 \over m_b^2}$ describes the falloff of the ground state, and we have approximated the integration limits to the desired accuracy. At the lower limit of the integral $t=(\delta m_Q - \Delta)^2$, this falloff attains,
$- \rho^2 {2 \Delta \delta m_Q \over m_b^2}$ ; the integral is found to be of order $1/m_Q$ with respect to the partial rate {\it over the same region}, which we can evaluate at leading order by the corresponding free quark rate $\int_{(\delta m_Q - \Delta)^2}^{(\delta m_Q)^2} dt |\vec{q}{\,}|$ :
\bea
{\delta \Gamma_I \over \Gamma_{partial}} \simeq-{3 \over 5} {\rho^2  \over m_b^2} (2 \Delta \delta m_Q)=-{3 \over 5} { m_d \delta m_Q \over m_b^2}
\eea 
where we have used that $\rho^2 \Delta={m_d \over 2}$ in the H.O. model. The last expression is seen to be of order $1/m_Q$. Indeed, in the $1/m_Q$ expansion, $\delta m_Q$ must be counted as $(m_Q)^1$. 

However, this is {\it not} what is to be done to evaluate the degree of duality violation on the total decay rate ; as we have remarked in our paper \cite{OH}, the real violation of duality is in fact much smaller, because the term must not be divided by the decay rate over the above small region of phase space, but rather by the total decay rate,  i.e.  the one over a much larger region of phase space. We end with :
\bea
{\delta \Gamma_I \over \Gamma_{tot}} \simeq -{3 \over 5}\rho^2 {2 \Delta \delta m_Q  \over m_b^2}
\left({2 \Delta  \over \delta m_Q }\right)^{3/2}=-{3 \over 5} {m_d \delta m_Q  \over m_b^2}
\left({2 \Delta  \over \delta m_Q }\right)^{3/2}
\eea
Parametrically, this is suppressed with respect to $1/m_Q$, because of the factor $({2 \Delta \over \delta m_Q })^{3/2}$ ($\delta m_Q >>\Delta$). 

Effect II. In another example, relying on a model of two-body decay, the article \cite{isgur} tries to take into account also the effect due to the 
$m_d \delta m_Q \over m_b^2$ terms present in {\bf partial rates} ; such terms, which corresponds to $1/m_Q$, are present {\bf separately} in the various exclusive channels, for instance one has for the ratio $R_{sl}^{(ground~state)}$ of the ground state to the free quark decay rates :
\bea
R_{sl}^{(ground~state)}=1+{3 \over 2} {m_d \delta m_Q \over m_b^2} +...
\eea 
but they cancel in the total decay rate. Then the argument would be that if the kinematical situation is 
such that only the ground state is produced, the total ratio $R_{sl}$ would depart from $1$ by the term  
${3 \over 2} m_d \delta m_Q \over m_b^2$. However, we notice that the effect is for
$t$ above the $D^{**}$ threshold $(m_B-m_{D^{**}})^2$, i.e. in a limited region of phase space. 
When we duely refer this effect to the total rate, we end with :
\bea
{\delta \Gamma_{II} \over \Gamma_{tot}} \simeq {3 \over 2} {m_d \delta m_Q \over m_b^2  }
{\int_{(\delta m_Q - \Delta)^2}^{(\delta m_Q)^2} dt |\vec{q}{\,}| \over \int_{0}^{(\delta m_Q)^2}dt |\vec{q}{\,}|} \simeq {3 \over 2} {m_d \delta m_Q \over m_b^2}
 \left({2 \Delta  \over \delta m_Q }\right)^{3/2}
\eea
which is once more parametrically smaller than $1/m_Q$, by the factor $({2 \Delta \over \delta m_Q })^{3/2}$. 

Finally, we conclude that both effects are not ${\cal O} (1/m_Q)$, but much smaller. In 
a $1/m_Q$ expansion with fixed ratio of heavy masses, the estimate of these terms,
often referred to as a potential $1/m_Q$ duality violation should be actually much smaller :
\bea
{\delta \Gamma_{I,II} \over \Gamma_{tot}}={\cal O}((1/m_Q)^{5/2})
\eea
Numerically too, sticking to the strict harmonic oscillator, and with realistic parameters, $m_b=5~GeV$, $m_c=2~GeV$, $m_d=0.3~GeV$ (constituent masses), $R^2=10~GeV^{-2}$, giving $\Delta=0.33~GeV$ (in agreement with heavy light spectroscopy), the effects are small : we find respectively : $ -2.2~ 10^{-3}$ and $5.6~10^{-3}$. Note that these parameters give a low $\rho^2$,
much lower than the one observed ($\rho^2 \simeq 0.5$ instead of $\rho^2 \simeq 1$)
a well known defect of the non relativistic approach ; it is then tempting, as done by some authors, to introduce additional "fudge" factors to account for a more realistic $\rho^2$ , which implies that the transition amplitudes will not be given by the basic Hamiltonian ; however such a procedure cannot be consistent and will lead to violation of the basic SV sum rules ; therefore, although presenting possibly some phenomenological advantage, it is misleading for the discussion of duality (in fact, the phenomenological discrepancy for the value of $\rho^2$ is corrected consistently in a relevant {\bf relativistic} approach \cite{wambach},\cite{BT}). Note also that we have taken the maximum effect
by cutting the integrals just at the $D^{**}$ threshold.
\par \vskip 1 truecm

\section{Complete cancellation of alleged terms} \label{cancellation} Having established the explicit expression of the controversial effects,
and demonstrated that they are much smaller than stated, one can further wonder whether such terms are really present at all in the actual departure from free-quark duality. Indeed, they have been picked out as isolated contributions to the decay rate, not as the full contribution at some definite order $(1/m_Q)^{5/2}$. The answer is then that {\bf these terms simply do not survive when one duely sums up all the contributions} of the same order to the total rate . Indeed, they correspond to non analytic, fractional powers of $(\delta m_Q)$, and such terms simply {\it do not appear} in the expansion, described above, eq. (\ref{ROH}), of the ratio $R_{sl}$ of the harmonic oscillator model, which derives from a complete and very explicit calculation. Moreover, the controversial terms are larger than the last term  calculated in eq. (\ref{ROH}) : they are $(1/m_b)^2 1/(\delta m_Q)^{1/2}$, while we have calculated to $(1/m_b)^2 1/(\delta m_Q)$ included. Therefore, they really cannot be there, they will be cancelled with similar terms when duely making the complete calculation of the total rate, as we have done. That such a cancellation can occur is already suggested by our calculations of section \ref{real}, which shows that the two contributions invoked by ref. \cite{isgur}  are of the same type $ \propto {m_d \delta m_Q \over m_b^2} ({ \Delta \over \delta m_Q })^{3/2}$, but of opposite sign ; of course, the cancellation is not complete at this stage, but it simply means that, in the approach of \cite{isgur}, one has still overlooked some similar effects, which is difficult to avoid since one just points out some possible contributions ; one does not rely on a systematic enumeration, as we try to do in our expansion. 
\par \vskip 1 truecm

\section{OPE versus exact numerical calculation at finite masses} \label{numerical} Finally, to discard definitively the worry about a possible large (i.e. several $10^{-2}$) departure from OPE, one can ask what is the departure of OPE from the {\bf exact} rate for physical masses and {\bf realistic parameters} in the model, i.e. far from the S-V limit. An exact numerical calculation can be made in the case of the H.O. model. We find that, for this H.O. model, close to the one
used in \cite{isgur}, but without fudge factors or adjustment of phase space, and in an {\bf exact} calculation where we sum  over all the eight allowed H.O. levels,
and with {\bf realistic parameters}, $R_{sl}$ is {\bf exceedingly close} to the OPE prediction. With the above parameters, the departure from 1 is:
\bea
R_{sl}-1 = -1.33~10^{-3} \label{num}.
\eea
This already small number has the same order of magnitude as the one predicted by the first two terms in the S-V expansion, eq. (\ref{ROH}), but the sign is {\it opposite}, which suggests the importance of the further terms in the S-V expansion ; this is not surprising since $m_b-m_c$ is not sufficiently small to make this expansion very effective : there is no reason in that case to reach the correct result with a small number of terms, and one would require a calculation of many terms. 

Therefore, for simplicity, instead of the S-V expansion, we now pass to the full $1/m_Q$ expansion, which we have indeed also performed. Let us recall that in this new OPE expansion, it is not assumed that $m_b-m_c$ is small with respect to $m_b$; each term of this new expansion includes a series of many S-V terms. As we have said, we have no term of order $1/m_Q$. Considering the $1/m_Q^2$ term in this new expansion, the above discrepancy of sign disappears : the OPE estimate of $R_{sl}-1$ is now negative and small. The correct magnitude itself is reproduced by including also the $1/m_Q^3$ term (to the accuracy $10^{-5}$). The $1/m_Q^4$ term is very small.
Let us insist that in the model, the coefficients of this $1/m_Q$ series have been {\bf demonstrated} to be the ones predicted by the OPE (in fact, the short time expansion) , so it means that the possible discrepancy with the OPE series is of magnitude $10^{-5}$ (probably at worst) for realistic masses.

The fact that the $1/m_Q^4$ is very small, while the $1/m_Q^3$ term is of the same order as the $1/m_Q^2$  one, is exactly as expected from the arguments of the paper
\cite{benson/hep-ph/0302262} in QCD : in fact, it is the $1/m_Q^2$ term which is abnormally suppressed, whence the abnormally small decrease from $1/m_Q^2$ to $1/m_Q^3$; after that, the terms are decreasing quite rapidly, as expected from the effect of inverse powers of $m_b-m_c$.

In summary, in this well-defined and completely calculable quark model, we retrieve 
in an explicit and exact calculation, without any assumption, the conclusion that have been inferred in QCD 
on the OPE of inclusive semi-leptonic decays, and which had been controversed
on the basis of certain misinterpreted previous quark model calculations .
\vskip 0.5cm
{\bf Acknowledgments} 

D.Melikhov acknowledges financial support from the Alexander von Humboldt-Stiftung. 
A.L.-Y., L. O., O.P. and J.-C.R. thank very much P. Roudeau and A. Stocchi 
for their continuous interest and support on the subject. 
They have also have benefitted from very useful discussions with N. Uraltsev 
on duality in the past years. They are grateful to Paolo Gambino, Achille Stocchi and Laurent Lellouch for their efforts in the edition
of the corresponding part of the Yellow Book.
They acknowledge partial support from the EC contract HPRN-CT-2002-00311 (EURIDICE).

\end{document}